\documentclass[final,3p,times,twocolumn]{elsarticle-2}

 \usepackage{graphicx}

\usepackage{amssymb}

\begin{document}
\begin{frontmatter}
%

\title{Defect-Induced Magnetism in Solids}

\author[l1]{P. Esquinazi \fnref{label2}}
\author[l2]{W. Hergert}
\author[l1]{D. Spemann}
\author[l1]{A. Setzer}
\author[l3]{A. Ernst}
\address[l1]{Institute
for Experimental Physics II, University of Leipzig,
Linn\'{e}stra{\ss}e 5, D-04103 Leipzig, Germany}
\address[l2]{institute of Physics,
Martin Luther University Halle-Wittenberg, 06120 Halle, Germany}
\address[l3]{Max Planck Institute of
Microstructure Physics, 06120 Halle, Germany}
\fntext[label2]{Corresponding author. Tel/Fax: +49 341 9732751/69.
E-mail address: esquin@physik.uni-leipzig.de (P. Esquinazi)}


%






\begin{abstract}In the last years the
number of nominally  non-magnetic solids showing magnetic order
induced by some kind of defects has increased continuously. From
the single element material graphite to several covalently bonded
non-magnetic compounds, the influence of defects like vacancies
and/or non-magnetic ad-atoms on triggering magnetic order has
attracted the interest of experimentalists and  theoreticians. We
review and discuss the main theoretical approach as well as
recently obtained experimental evidence based on different
experimental methods that supports the existence of defect-induced
magnetism (DIM) in non-magnetic as well as in  magnetic materials.
\end{abstract}


\end{frontmatter}




%


\section{Why Defect-Induced Magnetism was recognized so late?}
%
%
%
%

In the original publication of Heisenberg,
published in Leipzig in 1928 \cite{heisenberg28}, about the basic
concepts on the origin of  magnetic order in solids, it is written
at the end of the paper that the principal quantum number of the
electrons responsible for the magnetism must be $n   \gtrsim 3$.
It took several decades till scientists started thinking that
magnetic order may exist beyond this $n   \gtrsim 3$ condition.
The reason why magnetism based on $s$ and $p$ electrons was
finally predicted, discovered and  recognized so late, is mainly
due to a mixture of three things, namely: The first reason is
theoretical, since  Heisenberg's successful theory on magnetic
order  promoted a kind of magnetic prejudice against magnetic
signals coming from compounds with full $d$- or $f$-bands or
materials with only $s$ and $p$ electrons. Second, the
contribution of magnetic impurities and their usually difficult
characterization put also hard constraints, which till now are not
always removed.
Finally, the phenomenon of defect-induced magnetism (DIM)   in
systems without usual magnetic ions is based on the effect of
different kinds of defects upon the system. The production of
samples with a homogeneous distribution  of defects at the  right
lattice positions remains difficult.
Therefore, the obtained magnetic signals  are in some cases   so
small, that even tens of ppm of magnetic Fe would be enough to
produce similar ones.
Thus, the answer to the question of this section is a mixture of
technical  capabilities to check for the impurity contribution,
scientific (over)skepticism and the nature of the DIM phenomenon
itself.

In spite of these ``difficulties", the DIM phenomenon has  been
finally observed  in a broad spectrum of materials, from
carbon-based to several oxides  and the obtained evidence of the
last ten years leaves little doubt about its existence. In this
contribution we do not try to review the  huge amount of studies
published but we would like to emphasize a few new theoretical and
experimental results from different groups as well as from us
obtained in the last years that we believe should be of interest
for all scientists working on this subject.

\section{Basic ideas and general theoretical approach to the problem}

In parallel to the experimental exploration  of the unusual
magnetism, not based on $d$ or $f$ electrons, theoretical studies
emerged. Theoretical investigations can be based on model
Hamiltonians to study  basic physical features of the problem
\cite{dietl2000zener,yaz10}.  Detailed considerations of native
defects, hydrogen or light impurities in carbon or oxides require
calculations on the \textit{ab initio} level, based on density
functional theory (DFT). The application of  sophisticated
computer codes is not without problems and the complete discussion
of DIM demands a multicode approach.

An intensive search for new routes to  ferromagnetic oxidic
materials started with the investigation of Elfimov \textit{et
al.} \cite{elfimov2002possible}. Using CaO as an example, it was
demonstrated that dilute divalent cation vacancies in oxides with
rocksalt structure lead to a ferromagnetic ground state.  The
quite general result in \cite{elfimov2002possible}  that the spin
triplet state is the ground state, is also applicable to other
compounds with vacancies in octahedral coordination. In terms of
this somehow initiating paper an innumerable series of papers
appeared during the last 10 years investigating the electronic and
magnetic properties of vacancies and non-magnetic impurities in
different oxides, whereas especially MgO and ZnO attracted the
attention
\cite{wu2010magnetism,slipukhina2011ferromagnetic,mavropoulos2009ferromagnetism}.

A consistent theoretical proof of a  stable ferromagnetic ground
state consists of several steps. First, one has to calculate the
magnetic properties of the corresponding vacancies and impurities
in the dilute limit. Second, the mechanism of magnetic interaction
as a requirement of long-range magnetic order has to be
investigated. Third, a calculation of the transition temperature
has to predict the temperature range of ferromagnetic order. All
steps are connected with problems, as discussed by Zunger
\textit{et al.} \cite{zunger2010quest}. Those problems and the
restricted knowledge of the nature and distribution of defects in
experimental investigations, usually used as input of
calculations, limit the predictive power of \textit{ab initio}
calculations.

As a result of DFT  calculations cation vacancies in ZnO carry a
magnetic moment of $1.89 \mu_B$ \cite{adeagbo2010zno}. The
magnetic interaction of such defects breaks down if localization
corrections \cite{zunger2010quest} are taken into account.
Calculations on Open-shell impurity molecules like C$_2$ in ZnO
 seem to be a possible way for long-range
ferromagnetic order in ZnO  as calculations demonstrated
\cite{wu2010magnetism}.

Surfaces provide another route to controlled  ferromagnetism in an
otherwise non-magnetic host material. We studied room-temperature
$p$-induced surface ferromagnetism at the oxygen-terminated
ZnO(001) surface \cite{fischer2011room}.  The pseudopotential code
SIESTA was used to relax the structure. For a more adequate
description of the correlated electrons a multiple  scattering
Korringa-Kohn-Rostoker code \cite{luders2005self} was used. The
code provides the real space exchange coupling constants, which
serve as an input to a Monte Carlo simulation in the framework of
a classical Heisenberg model to determine the Curie temperature.
Finally it was shown, that the surface is thermodynamically stable
and ferromagnetic at room temperature \cite{fischer2011room}.

\section{DIM in carbon}

Already in 1968 the possibility to have magnetic order in
hypothetical hydrocarbons was emphasized by Mataga in a short
paper \cite{mataga}. Within the same line, in 1974 Tyutyulkov and
Bangov proposed theoretically the existence of unpaired electrons
in hydrocarbon molecules and nonclassical $\pi$-conjugated
polymers \cite{tyu74}. Ovchinnikov and coworkers followed a
similar line \cite{ovchi78,ovchi88} and in 1991 they proposed a
pure carbon structure based on 50\% graphite and 50\% diamond
bondings  that could show magnetic order with a saturation
magnetization above 200~emu/g \cite{ovchi91}, comparable to
$\alpha$-Fe. Although this structure apparently was not reproduced
or realized, later experiments with disordered or amorphous carbon
obtained by pyrolysis of certain precursors indicated the
existence of magnetic order with critical temperature above 500~K
and saturation magnetization $\sim 10~$emu/g at 4.3~K
\cite{murata91}. Further details of this early work can be seen in
\cite{esqui06}. The main problem of the early work is the unclear
contribution of magnetic impurities, basically due to insufficient
characterization of their density and reproducibility of the
reported phenomenon. In this section we discuss results related to
the magnetic order mainly found and confirmed in graphite samples.
Recent experimental studies  did not  provide clear evidence for
the existence of this phenomenon in single layer graphene
\cite{ric11,nai12} and therefore we will not discuss them here due
to the available space. The importance of the 3D lattice structure
of graphite to trigger magnetic order through carbon vacancies or
bonded hydrogen has been emphasized in  \cite{yaz08}, see also
\cite{yaz10}.

\begin{figure}[]
\includegraphics[width=3.2in]{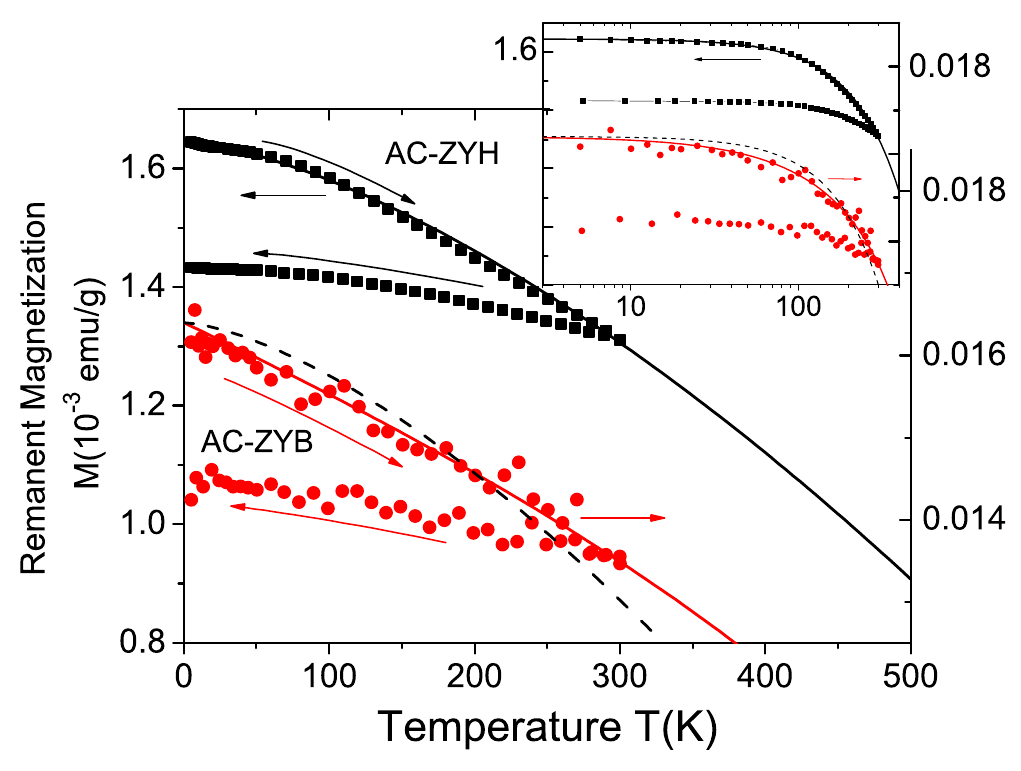}
\caption{Remanent magnetization at zero field measured after field
cooling the samples in a field of 1~T to 5~K. The data have been
taken by warming to 300~K and cooling down to 5~K, see arrows. The
results of two HOPG samples are shown: AC-ZYH (upper black data
points) and AC-ZYB (lower red data points). The inset shows the
same data but in a semi logarithmic scale. The continuous black
top line was calculated using the 3D Bloch $T^{3/2}$ model with
the following parameters: Curie temperature $T_c = 800~$K, a ratio
$2JS/k_B = 210~$K ($J$ the exchange coupling and $S$ the total
spin). For the AC-ZYB sample the remanence follows an anisotropic
2D spin waves model, continuous red line \cite{lev92,ser93}, which
follows a nearly linear decrease of the magnetization with
temperature. The continuous line was calculated using the
following parameters: Critical temperature $T_c = 600~$K,
spin-wave critical temperature due to low-energy spin-wave
excitations $T_c^{SW} = 1950~$K and anisotropy $\Delta = 0.001$.
For comparison we  show the theoretical (dashed) line obtained
within the 3D Bloch $T^{3/2}$ model with the best parameters set
to fit the data of sample AC-ZYB.} \label{gra}
\end{figure}

\subsection{DIM in as-received graphite samples and the
contribution of magnetic impurities: A never end story}

\begin{figure}[]
\hspace{-0.5cm}
\includegraphics[width=3.3in]{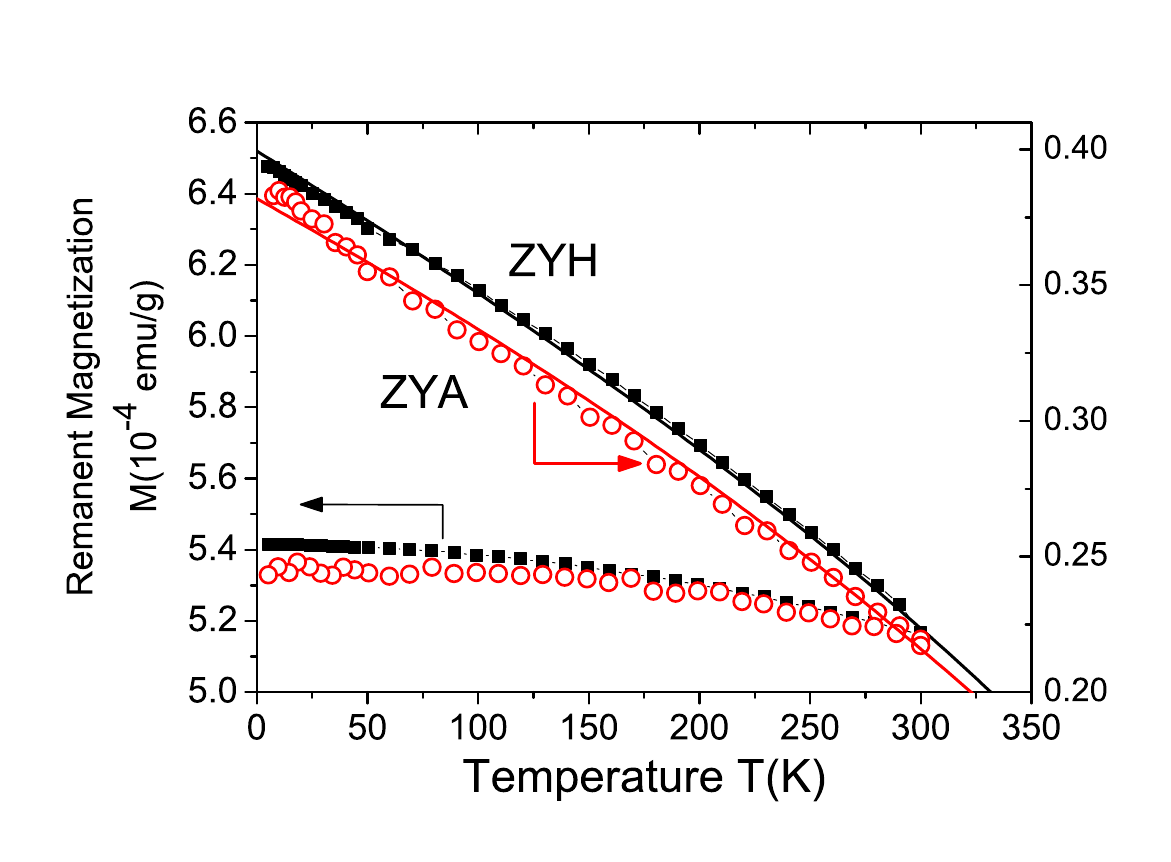}
\caption{Temperature dependence of the remanent magnetization,
similar as in Fig.~\ref{gra}, for two HOPG samples after field
cooling the samples in a field of 1~T to 5~K. The continuous lines
follow the 2D Heisenberg model with anisotropy \cite{lev92,ser93}
and were obtained with the  following parameters for sample ZYH
(ZYA): Critical temperature $T_c = 550~$K (750~K), spin-wave
critical temperature due to low-energy spin-wave excitations
$T_c^{SW} = 1700~$K (830~K) and anisotropy $\Delta = 0.001$.}
\label{gra2}
\end{figure}

The relatively small ferromagnetic moments obtained in most of the
materials that show DIM due to the actually ``brute force"
production methods used to trigger this phenomenon, make the knowledge
of  the ferromagnetic contribution from magnetic impurities
unavoidable to  assure that the measured phenomenon is intrinsic.
In this section we discuss mainly the magnetic contribution of
impurities in the as-received state of the samples.
Whatever is the as-received state of the sample to be characterized,
one should not forget that ultrasonic cleaning is nevertheless
required to get rid of the edge contamination graphite samples may have.
The influence
of defects or  ion irradiation  to the magnetism
of graphite will be discussed in the next section.

There are basically two main problems to characterize the magnetic
response from  impurities using only the SQUID
magnetometers. Firstly, the measurement of their concentrations.
For concentration below 50~ppm there are not too many experimental
methods able to measure this content with enough certainty.
Second, the knowledge of their concentration alone is not enough.
One needs to know at least the typical size of the impurity grains
inside the material of interest,  which requires a method for elemental imaging that provides excellent
detection limits in the ppm and sub-ppm range together with reliable quantification, preferably
in a non-destructive way. The sole measurement of the
magnetic moment of a sample with a known amount of impurities does
not provide always with a clear statement whether the
ferromagnetism is or is not due to impurities. In this section we
provide a simple example of this problem using the magnetization
data of four as-received HOPG samples of different origins and
impurity concentrations. A complete description of all
measurements done in as-received HOPG samples including the
elemental analysis as a function of position inside the samples
using particle induced x-ray emission (PIXE, see for example
\cite{jems08}) will be published elsewhere.

Figure \ref{gra} shows the temperature hysteresis of the remanent
(zero field) magnetization after cooling the samples in a field of
1~T, as a function of temperature of two HOPG samples with Fe
concentration (the main magnetic impurity) $23~\mu$g/g ($\simeq
5.8$~ppm) for sample AC-ZYH and $0.2~\mu$g/g ($\simeq 0.05$~ppm)
for sample AC-ZYB, within a relative error  $\lesssim 10$\%. As
shown in \cite{jems08} the Fe is mainly distributed in spot-like
regions of diameters  $\lesssim 10~\mu$m all over the sample
interior (penetration depth of the PIXE analysis $\gtrsim
30~\mu$m). For the sample AC-ZYH one recognizes a larger density
of Fe atoms in the spots than in the spots of sample AC-ZYB, both
samples appear to have similar  density of spots. Therefore, the
difference in Fe concentration of a factor 115 between the two
samples is  probably due to the difference in Fe concentration
within the spots and also a bit due to spot size, because the
spots (grains) appear  larger for the  AC-ZYH sample. Both samples
show the typical field hysteresis with a ratio between saturation
values at 5~K of $M_s($AC-ZYH)$/M($AC-ZYB)$\simeq 6 \times 10^{-3}
/ 4 \times 10^{-5} = 150$ (both values in emu/g). If {\em all} Fe
would be ferromagnetic,  from the ratio in concentration we would
conclude that the Fe concentration roughly explains the difference
in magnetization at saturation values as well as the ratio of 102
in the remanent magnetization, see Fig.~\ref{gra}. Taking into
account that $1~\mu$g/g of ferromagnetic Fe (Fe$_3$O$_4$) in
graphite would produce a magnetization at saturation of $2.2
\times 10^{-4}~$emu/g ($1.4 \times 10^{-4}~$emu/g), if
\textit{all} the measured Fe concentration would be ferromagnetic
we would have the magnetization values at saturation of $5 \times
10^{-3}~$emu/g ($3.2 \times 10^{-3}$~emu/g) for sample AC-ZYH and
$4.4 \times 10^{-5}~$emu/g ($2.8 \times 10^{-5}$~emu/g) for sample
AC-ZYB. From these estimated values it appears that all the
measured Fe, if ferromagnetic, would be enough to explain the
observations. It should be noted, however, that Fe in general is
not homogeneously distributed in the micron-sized impurity grains
as revealed by PIXE elemental imaging. For example, the main
impurity contamination in these grains in AC-ZYB is Ti and V, but
in AC-ZYH the Fe and V concentrations are similar whereas the Ti
contamination can be neglected. Consequently, the impurity grains
in both samples are different and cannot be considered to be of
pure Fe (or Fe$_3$O$_4$).

Furthermore, Fe in graphite not always shows a ferromagnetic
behavior or induces one, upon grain size and distribution. For
example, in \cite{pabloprb02} a sample with an inhomogeneous Fe
concentration of up to 0.38\% (in weight) shows no magnetic order
at all. Moreover, no increase in the magnetic order, existing in
the as-received state, was measured in HOPG samples after
implanting Fe up to concentrations of $4000~\mu$g/g
\cite{hoh08}. In this last case, the obtained result
appears reasonable because after implantation the Fe atoms reside
as single atoms randomly distributed in the (disordered) graphite
lattice. The $T-$dependence of the remanent magnetization suggests
that not only Fe ferromagnetism is at work in the as-received
samples.  As shown in \cite{jems08},  to compare quantitatively
the $T$-dependence of the ferromagnetic signal with appropriate models
avoiding arbitrary background subtractions, we take the remanent
magnetization measured  at zero applied field.  The
observed hysteresis between warming and cooling, see
Fig.~\ref{gra}, is a clear evidence for the existence of a
ferromagnetic state with Curie temperature above 300~K.

For large enough 3D ferromagnetic Fe particles we expect to see a
$T-$dependence of the remanent magnetization given by the
excitation of spin waves following the usual 3D Bloch $T^{3/2}$
model \cite{kitt}. The sample with the largest Fe
concentration (AC-ZYH) shows a $T-$dependence for the remanent
magnetization compatible with this law, see Fig.~\ref{gra}.
However, for the other sample with 0.05~ppm Fe concentration, the
$T-$dependence deviates from this law but decreases quasi-linearly
with $T$, the same dependence observed  for the magnetization of
irradiated HOPG samples and interpreted in terms of 2D Heisenberg
model with a weak anisotropy \cite{barzola2,xia08}, see
Fig.~\ref{gra}. We may
conclude that for the sample with the smaller Fe
concentration a non-negligible part of the observed magnetic order
comes from defects that induce a two dimensional anisotropic
magnetism and is not  due to ferromagnetic Fe or other
magnetic impurities.

To check the above correlation  we did similar measurements in two
other samples of different origin. Figure~\ref{gra2} shows the
results for the samples ZYH and ZYA with Fe concentrations (main
impurity): $10.2~\mu$g/g
 and $0.55~\mu$g/g. The
ratio between magnetizations at saturation at 5~K is
$M_s($ZYH)$/M_s($ZYA)$\simeq 2.3 \times 10^{-3} / 2.8 \times
10^{-4} = 8.2$ (values in emu/g).  This ratio does not follow the
expected ratio (19)  if {\em all} Fe would contribute to the
ferromagnetic signal. Whereas $M_s($ZYH) appears to be
 compatible assuming that {\em all} Fe concentration
would contribute ferromagnetically, $M_s($ZYA) is a factor of 2.3
larger than the highest expected saturation magnetization. From
this we conclude that an extra contribution produces the observed
magnetic order. Interestingly, both samples show a quasi-linear
temperature  dependence for the remanence magnetization, see
Fig.~\ref{gra2}, which can be well fitted within the 2D Heisenberg
anisotropic model. From all these studies we may conclude that
extra contributions, other than those from magnetic impurities, to
the observed ferromagnetic magnetization response exist in
as-received graphite samples. No general answer can be given,
however, even knowing the magnetic impurity concentration, to the
question whether magnetic impurities are or are not the reason for
the observed magnetic response in a given sample. For samples with
a relatively large amount of magnetic impurities \cite{sep12} it
has little sense to speculate whether the DIM phenomenon can be
clearly observed from the measurements.

\subsection{The role of vacancies and hydrogen}

The existence of DIM in graphite, as-received samples of different
magnetic impurity contents \cite{yakovjltp00,pabloprb02} as well
as after inducing defects by ion irradiation \cite{pabloprl03},
was later confirmed  by independently done studies
\cite{ohldagl,zha07,xia08,yan09,ohldagnjp,uge10,yaz10,ram10,he11,mia12}.
The main idea to interpret the existence of magnetic
order in graphite \cite{yaz08,pis07,yaz07} (for
 reviews see \cite{yaz10,vol10}) is based in the long range
interaction that appears between the nearly localized magnetic
moments existing at carbon vacancies \cite{lehtinen04,ma04},
 or when a proton is bonded to
a carbon $p$-electron normal to the graphene layer
\cite{duplock04}. That single vacancies in graphite can trigger a
magnetic moment has been proved experimentally by STM spectroscopy
\cite{uge10} at the surface of a bulk graphite sample as well as
by SQUID measurements of bulk samples irradiated with different
ions and doses \cite{ram10}. As expected, magnetic order was found
only in samples in which a vacancy density of several percents was achieved, i.e.
 a distance between them of the order of 2~nm \cite{ram10}. This density is
necessary to get magnetic order in solids,
independently of the details of the structure or elements that are
in the lattice, provided that the vacancies or other defects in
the lattice lead to nearly localized magnetic moments. Further
recent studies on the magnetic order in graphite triggered by proton and helium irradiation
were done in \cite{mak11};  the observed magnetic order  appeared to be
linked to defects in the graphite planes, like vacancies.  Electron spin resonance
studies on proton-irradiated
HOPG samples at different fluences indicated
the existence of metalliclike islands surrounded by insulatinglike
magnetic regions \cite{lee10} in agreement with previous findings
\cite{sch08}.

The intrinsic origin of the magnetic order triggered by proton irradiation on graphite has
been backed by transmission x-ray magnetic circular dichroism (XMCD) studies
\cite{ohldagl}. That study left no doubt that carbon can be
magnetic without the need of magnetic impurities. Further  XMCD studies in
as-received as well as in proton irradiated HOPG samples \cite{ohldagnjp}
showed that not only  the carbon $\pi$-band is spin polarized but
hydrogen-mediated electronic states
also exhibit a net spin polarization with significant magnetic remanence at
room temperature \cite{ohldagnjp}.  The obtained results
showed that the magnetic signals originated mostly from a
$\sim 10~$nm  near-surface region of the sample, where the
saturation magnetization may reach up to 25\% of that of Ni. The
results also indicated that hydrogen plays a role in the magnetic
order but it is not implanted by the irradiation but should come
from dissociation of H$_2$ molecules at the near surface region of
the HOPG sample \cite{ohldagnjp}. These  XMCD results support the
findings from a low-energy muon spin rotation experiment on HOPG
samples that indicated the existence of a ferromagnetic surface of
$\sim 15~$nm thickness \cite{dubman:08}. Further theoretical work
showed that the magnetic coupling becomes weaker when the
hydrogen-hydrogen distance increases \cite{pei06,pis07}.

According to \cite{duplock04} hydrogen absorption on a graphene
sheet as well as hydrogen chemisorption in graphite \cite{yaz08},
may lead to the formation of a spin-polarized band at the Fermi
level and robust ferromagnetic order should appear. These
theoretical studies are supported by the XMCD results referred
above \cite{ohldagnjp} and emphasize the need for further studies
on the role of hydrogen in the magnetism of graphite. Searching
for a simple method to trigger magnetic order in graphite samples
of mesoscopic size through hydrogen doping, the authors in
\cite{barapl11} treated graphite surfaces with sulfuric acid. It
is known that this kind of acid treatment leads to hydrogen doping
in the graphite structure. Indeed, the magnetization measurements
of micrometer small graphite grains treated with sulphuric acid
showed clear signs for magnetic order, which amount depends on the
used dilution of the acid as well as on the treatment time; it
decreased after mild annealing in vacuum \cite{barapl11}. Further
evidence for the existence of magnetic order triggered by the acid
treatment came from the anisotropic magnetoresistance (AMR),
defined as the dependence of the resistance on the angle between
the direction of the electric current and the magnetic field, both
applied parallel to the main area of the sample \cite{barapl11}.
The reported results indicated that the $L-S$ coupling in graphite
is not negligible when a magnetic moment is originated by hydrogen
doping (or due to vacancies). The observed rather large AMR values
support a hydrogen-mediated magnetism in graphite in agreement
with the XMCD results of \cite{ohldagnjp}.


\section{Evidence for DIM in Oxides}

Nearly simultaneously with reports on magnetic  order in graphite
about 12 years ago, the search for ferromagnetism in diluted
magnetic semiconductors attracted the interest of a broad
community. This was basically due to the expectations of combining
the advantages of semiconductors into spintronics applications,
for example. However, the early excitement after the first reports
on magnetic order at room temperature appeared, was quickly
overwhelmed by doubts on homogeneity issues as well as extra
contaminations. On the other hand the broad research done
afterwards helped to recognize that, as in graphite, defects, as
vacancies (or added nonmagnetic ions) play a crucial role in the
observed magnetic order. For recent reviews on DIM in oxides the
reader should refer to  \cite{ogale10,sto10,vol10}. Here we
restrict ourselves to point out some results regarding DIM due to
vacancies, hydrogen and surface states in certain oxides reported
recently.

The ground state of cation vacancies (0,V,F centres) in oxides
attracted attention already in the 60's and 70's and there are
extensive studies of cation vacancies in simple oxides like
Al$_2$O$_3$, MgO, SrO, CaO, BeO and ZnO, for a review see, e.g.,
\cite{sch06}. A high spin state of the neutral Mg vacancy in MgO
was reported in \cite{hal73}, but probably the first observation
of a high spin state due to a cation vacancy was reported in a ZnO
sample, i.e. due to a Zn vacancy, treated by electron irradiation
\cite{gal70}. In spite of that the possibility to have magnetic
order through a minimum amount of vacancies
 was not recognized at
that time.

For undoped ZnO, probably the first  hints on the possible role of
vacancies in the observed magnetic order were obtained in thin
films prepared by pulsed laser deposition (PLD) under partial
N$_2$ atmosphere \cite{Xu08}. This rather preliminary result  was
confirmed a year later in \cite{Kha09}, a study that concluded
that neutral Zn vacancies, not O vacancies, produced during the
preparation of the film in the PLD chamber should play the main
role in the observed magnetic order. Characterization of the
lattice defects by x-ray absorption spectroscopy (XANES) at the Zn
K-edge in ferromagnetic, pure ZnO films, supported this conclusion
\cite{hau11}. We note that the absolute value of the magnetization
of these ferromagnetic thin films ($\sim 10^{-2}~$emu/g) suggests
already that the amount of ferromagnetic mass in the films is
certainly less than 1\%, an indication of the inhomogeneous
distribution of defects. Therefore one may still doubt whether
bulk characterizations of the films would provide the properties
of the magnetic regions. New  studies of un-doped ZnO films
prepared on silicon and quartz substrates suggested, however, that
the ferromagnetism is originated from singly occupied oxygen
vacancies, not Zn vacancies, reaching
 magnetization values of the order of 1~emu/g for $\sim 100~$nm thick films \cite{zha12}.
 The conclusion that oxygen vacancies in ZnO are
 the reason for the observed magnetic order is at odd with  several works cited above. However,  in that work
 \cite{zha12} no clear analysis of the magnetic impurities in the successive annealing
 steps was done. Therefore, the subject remains partially controversial.

The possibility of triggering magnetic order due to hydrogen
adsorption at the surface of pure ZnO  was studied theoretically
in \cite{San10,Liu09}. Evidence for surface  magnetism in pure ZnO
films after hydrogen annealing at $100^\circ - 500^\circ$ was
found in \cite{li11} together with evidence of the importance of
OH-terminated surfaces,   supporting theoretical predictions.
Interestingly, the FM could be turn on and off after annealing in
hydrogen or argon atmosphere. Further support to the possibility
of using hydrogen to trigger magnetic order in ZnO came from the
change in magnetization \cite{kha11} as well as in the
 magnetotransport properties \cite{kha12} after low energy proton
implantation on ZnO single crystals. The obtained magnetization at
saturation was $\sim 5$~emu/g and localized in a near surface
region of thickness $\lesssim 20~$nm \cite{kha11,kha12}.  The
measured anisotropic magnetoresistance (AMR) up to room
temperature indicates a spin splitted band as well as a finite
spin-orbit coupling \cite{kha12}.   These works indicate that
hydrogen should be a good candidate to trigger magnetic order in a
more systematic way than with solely vacancies. If the effect is
reproducible, triggering magnetic order through
 hydrogenation of the surfaces of micro- and nanowires of ZnO should be possible.

Not only in ZnO but in several other  oxide structures  like MgO,
SrTiO$_3$, MgAl$_2$O$_4$, LaAlO$_3$, their surface and upon
termination can show magnetic order at room temperature without
extra doping. This is the conclusion that was arrived in
\cite{kha10}. In particular the sensitivity of the magnetic
signals after acetone or ethanol cleaning of the surface of
SrTiO$_3$ substrates indicate a surface origin. The impact of
these two liquids on the surface magnetism has been theoretically
studied in \cite{ade11}. Those  results indicate that Ti- as well
as O-vacancies at the surface play a role in the observed
difference between ethanol and acetone influence on the surface
magnetism.

The studies in \cite{her10} showed that the observed room
temperature magnetic order in ZnO:Cu(2\%) films can be attributed
the magnetic moments arising at the Cu ion, i.e. the $d^{10}$
electronic  state of Cu can decrease when coupled to an O-vacancy
originating a finite moment coming from the Cu $d-$band; the
magnetic moments of O are found opposite oriented to the Cu ones.
Actually,  a similar explanation was proposed earlier
 in \cite{duha05} to explain  the ferromagnetism observed in TiO$_2$:Cu. All these results indicate the
important role that vacancies may play after doping non magnetic oxides to trigger magnetic order.
Not only vacancies but also dislocations appear to provide a main contribution to the magnetic response of undoped and Mn-doped ZrO$_2$ films \cite{zip10}.

The role of defects in the  magnetic response of oxides goes
beyond the nominally nonmagnetic oxides but can be also shown to
play a role in magnetic oxides stressing the fact that  DIM is a
quite general phenomenon. Recent experiments
\cite{torres2011evidence} provided evidence for DIM in
ZnFe$_2$O$_4$ samples grown under low O$_2$ pressure, pointing at
the role of oxygen vacancies in
 the observed magnetism.  This is further
 supported by recently done XMCD measurements and
 \textit{ab initio} calculations in the framework explained above \cite{nayak2013ferrite}.
All this work demonstrates that  a missing oxygen atom between two
adjacent Fe$^{3+}$ atoms on B sites leads to a parallel alignment
of the Fe moments and therefore a large magnetic moment per unit
cell.

We would like to end this section referring to photoconductivity
studies, a property that was hardly used in the past to
characterize the effects of DIM. The idea is to study the effects
of a magnetic field on the photoresistance, a property that
depends on the lifetime of photogenerated electrons and this on
the particularities of the energy centers  inside the gap that are
originated by defects. If these defects play a role in DIM, then a
magnetic field can  influence the photoconductivity. Measurements
in magnetic ZnO films  \cite{zap11} revealed that a magnetic field
enhances the recombination time of photoexcited carriers,
increasing the photoconductivity. In principle this property may
be used in the future as a magnetic defect spectroscopy, studying
the effect of a magnetic field on the photoconductivity in a broad
energy range.

\section{Evidence for DIM in other compounds}

We note here  two different non-oxide materials that after
irradiation show a ferromagnetic response. Proton irradiation of
MoS$_2$ revealed magnetic ordering at room temperature when
exposed to a 2~MeV proton beam (similar energy than in
\cite{pabloprl03}). The temperature dependence of magnetization
displays ferrimagnetic behavior with an remarkably high Curie
temperature of 895~K \cite{mat12}. The authors suggest that not
necessarily a single kind of defect but the combination of
magnetic moments arising from different defects, like vacancies,
interstitials, deformation and partial destruction of the lattice
structure may be necessary to understand the triggered magnetic
order. DIM was also observed after  neutron irradiation of SiC
single crystals \cite{liu11}. The authors in that work
demonstrated that mainly the produced divacancies (V$_{\rm
Si}$V$_{\rm C}$) appear to be responsible for the observed
magnetism. Theoretical studies revealed that extended tails of the
defect wave functions induce the long-range coupling between the
localized moments caused by the divacancies \cite{liu11}, a
further example of the richness of the DIM phenomenon in solids.

\section{Perspectives and Conclusion}

Maximum magnetization can be achieved for a certain vacancy
concentration, beyond it, the magnetization will reduce and
eventually  vanishes. This limit provides maximum magnetization
values that would hardly surpass that of usual strong
ferromagnets.  Therefore, the production of large-mass homogeneous
magnetic samples with DIM will remain difficult and future
activities should concentrate on inducing this phenomenon in
rather small samples. Although vacancies will remain an important
defect to take care, several studies discussed here (see also
\cite{vol10}) already indicate that doping with non-magnetic
elements as H, C, N, etc., appears nowadays an interesting route
to achieve high magnetization values and homogeneous samples. Due
to the relatively small values of magnetization one tends to
believe that DIM is a weak phenomenon, however they are small
because of the unidentified mass of the FM regions. Taking into
account that $\sim 5\%$ vacancies can trigger magnetic order with
$T_C
>300~$K, a comparison with $T_C \simeq 150~$K triggered by $\sim
5\%$ Fe-magnetic ions in Pd \cite{yeh75} indicates that DIM is not
a weak but an extraordinarily strong phenomenon.

\vspace{-0.3cm}
\section*{Acknowledgment}

This work was supported by the DFG within the
Collaborative Research Center (SFB 762) ''Functionality of
Oxide Interfaces."




\vspace{-0.3cm}
\bibliographystyle{IEEEtran}


%

\end{document}